\documentclass[journal,10pt,onecolumn,draftclsnofoot,]{IEEEtran}

\usepackage{cite}
\usepackage{textcomp}
\def\BibTeX{{\rm B\kern-.05em{\sc i\kern-.025em b}\kern-.08em
    T\kern-.1667em\lower.7ex\hbox{E}\kern-.125emX}}

\usepackage[utf8]{inputenc}
\usepackage[T1]{fontenc}
\usepackage{textcomp}
\usepackage{fixltx2e}
\usepackage{xcolor}
\usepackage{multirow}
\usepackage[pdftex]{graphicx}
\usepackage{epstopdf}
\AppendGraphicsExtensions{.tif}

\DeclareGraphicsExtensions{.pdf,.png,.jpg}
\usepackage[pdftex,bookmarks=false,colorlinks=false,pdfdisplaydoctitle=true,pdfborder={0 0 0}]{hyperref}

\makeatletter
\let\saved@hyper@linkurl\hyper@linkurl
\AtBeginDocument{
	\NoHyper
	\let\hyper@linkurl\saved@hyper@linkurl
}
\makeatother

\usepackage[font=scriptsize]{caption}
\usepackage[caption=false,font=footnotesize]{subfig}
\usepackage{booktabs}
\usepackage{url}
\usepackage{amsmath}
\usepackage{amsthm}
\usepackage[capitalise,noabbrev]{cleveref}
\usepackage{xspace}
\usepackage{booktabs}
\usepackage{epsfig}

\usepackage{enumitem}
\usepackage{url}
\usepackage{breakurl}

\usepackage[american]{babel}

\usepackage{acronym}
\acrodef{COVID-19}{coronavirus disease 2019}
\acrodef{RNA}{ribonucleic acid}
\acrodef{AI}{Artificial Intelligence}
\acrodef{CDC}{Centers for Disease Control and Prevention}
\acrodef{rRT-PCR}{real-time Reverse Transcription Polymerase Chain Reaction}
\acrodef{CT}{Computed Tomography}
\acrodef{CNN}{Convolutional Neural Network}
\acrodef{ML}{Machine Learning}
\acrodef{WHO}{World Health Organization}
\acrodef{NAAT}{Nucleic Acid Amplification Tests}
\acrodef{FDA}{Food and Drug Administration}
\acrodef{PCR}{Polymerase Chain Reaction}
\acrodef{DTL-MC}{Deep Transfer Learning-based Multi-Class}
\acrodef{CML-MC}{Classical Machine Learning-based Multi-Class}
\acrodef{DPC}{Dominant Principle Component}
\acrodef{PCA}{Principle Component Analysis}
\acrodef{DTL-BC}{Deep Transfer Learning-based Binary-Class}
\acrodef{MFCC}{Mel Frequency Cepstral Coefficients}
\acrodef{SVM}{Support Vector Machine}
\acrodef{FPR}{False Positive Rate}
\acrodef{FNR}{False Negative Rate}

\usepackage{tabularx}
\usepackage{array}
\newcolumntype{P}[1]{>{\centering\arraybackslash}p{#1}}
\usepackage{resizegather}

\DeclareRobustCommand*{\IEEEauthorrefmark}[1]{%
  \raisebox{0pt}[0pt][0pt]{\textsuperscript{\footnotesize #1}}%
}

\begin{document}

\title{AI4COVID-19: AI Enabled Preliminary Diagnosis for COVID-19 from Cough Samples via an App}

\author{{Ali Imran}\IEEEauthorrefmark{1,2},
	{Iryna Posokhova\IEEEauthorrefmark{2,3}, {Haneya N. Qureshi}\IEEEauthorrefmark{1}, {Usama Masood}\IEEEauthorrefmark{1}, {Muhammad Sajid Riaz}\IEEEauthorrefmark{1}, {Kamran Ali}\IEEEauthorrefmark{4}, \\ {Charles N. John}\IEEEauthorrefmark{1}, MD Iftikhar Hussain}\IEEEauthorrefmark{2,5} and {Muhammad Nabeel}\IEEEauthorrefmark{1} \\
\IEEEauthorrefmark{1}AI4Networks Research Center, Dept.\ of Electrical \& Computer Engineering, University of Oklahoma, USA \\
\IEEEauthorrefmark{2} AI4Lyf LLC, USA \\
\IEEEauthorrefmark{3} Kharkiv National Medical University, Ukraine \\ \IEEEauthorrefmark{4} Dept.\ of Computer Science \& Engineering, Michigan State University, USA \\
\IEEEauthorrefmark{5} Allergy, Asthma \& Immunology Center PC, USA \\
{Corresponding author: Muhammad Nabeel (e-mail: muhmd.nabeel@ou.edu)}
	}

\maketitle

\begin{abstract}
\textbf{Background:} The inability to test at scale has become humanity’s Achille’s heel in the ongoing war against the COVID-19 pandemic. A scalable screening tool would be a game changer.   Building on the prior work on cough-based diagnosis of respiratory diseases, we propose, develop and test an \ac{AI}-powered screening solution for COVID-19 infection that is deployable via a smartphone app. The app, named AI4COVID-19 records and sends three 3-second cough sounds to an AI engine running in the cloud, and returns a result within two minutes. 

\textbf{Methods:} Cough is a symptom of over thirty non-COVID-19 related medical conditions. This makes the diagnosis of a COVID-19 infection  by cough alone an extremely challenging multidisciplinary problem.  We address this problem by investigating the distinctness of pathomorphological alterations in the respiratory system induced by COVID-19 infection when compared to other respiratory infections. To overcome the COVID-19 cough training data shortage we exploit transfer learning. To reduce the misdiagnosis risk stemming from the complex dimensionality of the problem, we leverage a multi-pronged mediator centered risk-averse AI architecture. 

\textbf{Results:} Results show AI4COVID-19 can distinguish among COVID-19 coughs and several types of non-COVID-19 coughs. The accuracy is promising enough to encourage a large-scale  collection of labeled cough data to gauge the generalization capability of  AI4COVID-19.  AI4COVID-19 is not a clinical grade testing tool. Instead, it offers a screening tool deployable anytime, anywhere, by anyone. It can also be a clinical decision assistance tool used to  channel clinical-testing and treatment to those who need it the most, thereby saving more lives.

\end{abstract}


\acresetall

%
\section{Introduction}
\label{sec:section1}

By April 28, 2020, there were 3,024,059 confirmed cases of  \ac{COVID-19}, leading to 208,112 deaths and disrupting life in 213 countries and territories around the world~\cite{WHO}. The losses are compounding everyday.
Given no vaccination or cure exists as of now, minimizing the spread by timely testing the population and isolating the infected people is the only effective defense against the unprecedentedly contagious COVID-19.
However, the ability to deploy this defense strategy at this stage of pandemic hinges on a nation’s ability to timely test significant fractions of its population including those who are not contacting medical system yet.
The capability for agile, scalable and proactive testing has emerged as the key differentiator in some nations’ ability to cope and reverse the curve of the pandemic, and the lack of the same is the root cause of historic losses for others.

\subsection{Why might not clinic visit based COVID-19 testing mechanisms alone sufficiently control the pandemic at this stage?}
\label{sec:section1A}

The “Trace, Test and Treat” strategy succeeded in flattening the pandemic curve (e.g., in South Korea, China and Singapore) in its early stages. However, in many parts of the world the pandemic has already spread to an extent that this strategy is not proving effective anymore~\cite{bbc}.
Recent studies show that it is virus  often transmitted when an undiagnosed population coughs, that contributes to its much rapid and covert spread~\cite{cough_spreader}.
Data shows that 81\% of COVID-19 carriers do not develop severe enough symptoms  for them to seek medical help, and yet they act as active spreaders~\cite{cascella2020features}.
Others develop symptoms severe enough to prompt medical intervention only after several days of being infected.
These findings call for a new strategy centered on \textit{“Pre-screen/test proactively at population scale, self-isolate those tested positive for self-healing without further spreading and channel medical care towards the most vulnerable”.}

As per \ac{WHO} guidance, \ac{NAAT} such as \ac{rRT-PCR} should be used for routine confirmation of COVID-19 cases by detecting unique sequences of virus \ac{RNA}.
This test method, while being the current gold standard, is not an adequate way to control the pandemic for reasons that include but are not limited to:
\begin{enumerate}[wide,font=\itshape]
	\item The limited availability of testing due to geographical and temporal factors. 
	\item The scarcity and expense of
	clinical tests needed to cover the massive time-sensitive demand.
	\item The requirement of  in-person visits to a hospital, clinic, lab or mobile lab.
	Such  visits expose more members of the public to COVID-19.
	This is not a trivial problem given the recent studies that show how highly stable and hence contagious COVID-19 appears to be. For example,~\cite{van2020aerosol} shows that the aerosol stability of COVID-19 is up to three hours in aerosols and up to seven days on different surfaces.
	
	\item The turnaround time for current tests is several days, recently stretching to 10 days in some countries as labs are becoming overwhelmed~\cite{tendays,hospitals_overwhelm}.
	By the time a patient is diagnosed using current methods, the virus has already been passed to many.
	\item The in-person testing methods put the medical staff, particularly those with limited protection, at serious risk of infection. The inability to protect our medics can lead to further shortage of medical care and increased distress on the already stressed medical staff.
\end{enumerate}

To make tests more readily accessible, on March 28\textsuperscript{th} the United States \ac{FDA} approved a faster test that can yield results in 15 minutes~\cite{cnn2}.
The test works similar to \ac{PCR} by identifying a portion of the COVID-19 RNA in the nasopharyngeal or oropharyngeal swab. 
The FDA also recently approved another rapid molecular-based test, which delivers positive results in as little as five minutes and negative results in 13 minutes \cite{abott}. However, the FDA warns that there is a high  probability of false negative results using this test \cite{fda-warn}.
While a leap forward, this test still requires an office visit and thus the breaching of social distancing and self-isolation.
Though much faster, the newly approved test still does not solve many of the aforementioned problems.
Furthermore, emerging reports of shortages of critical equipment used to collect patient specimens, like masks and swabs, could blunt its impact on controlling the pandemic~\cite{masks_short1, masks_short2}. In order to protect others from potential exposure, the FDA has also approved at-home sample collection \cite{at-home}. However, once a patient collects a nasal sample, they need to put it in a saline solution and ship it overnight to a certified lab authorized to run specific tests on the kit. Hence, this approach also introduces delays and could compromise on the quality of samples if the sample is stored for too long. In addition, it could also introduce the chances of errors while collecting the sample, since the patients collect the sample themselves, rather than trained doctors or healthcare professionals.

More recently, two alternative approaches for COVID-19 infection diagnosis  leveraging  analysis of either X-ray~\cite{wang2020covid,XRAY1,XRAY2,XRAY3,XRAY_New1,XRAY_New2,XRAY_New3,XRAY_New4,XRAY_New5,XRAY_New6,XRAY-email-1, XRAY-email-2, XRAY-email-3} or CT Scan~\cite{xu2020deep,CT2,CT3,CT_New1,CT_New2,CT_New3,CT_New4} images have been proposed in the literature.
These techniques, either through  an examination by a radiologist, or when combined with AI-based image processing, are able to diagnose COVID-19 with even higher accuracies, and in some cases even better than the \ac{rRT-PCR} based test. Recent studies report a pooled sensitivity of 94\% (95\% confidence interval: 91\%-96\%), but a low specificity of 37\% (95\% confidence interval 26\%-50\%) for CT-based diagnosis \cite{ct-update-1}. Therefore, CT based diagnosis may help to overcome the sub-optimal sensitivity of PCR tests \cite{ct-update-2}.
However, while both of these approaches reduce the burden on radiologists to perform the diagnosis, they still require a visit to a well-equipped clinical facility.
As a result, these approaches also inherit the issues of office visit based tests that are highlighted above. 

\vspace{2.5mm}

\textit{It is mainly due to the inability to test large swaths of populations timely, safely and cost effectively and exactly track the actual spread that even the richest nations on earth are finding it difficult to contain the pandemic.}

\subsection{Proposed cough based COVID-19 screening approach}
\label{sec:section1B}

The idea of using cough for possible preliminary diagnosis of COVID-19, and the need to investigate its feasibility is motivated by the following key findings:

\begin{enumerate}[wide,font=\itshape]
	\item Prior studies have shown that cough from distinct respiratory syndromes have distinct latent features~\cite{thorpe2001acoustic,chatrzarrin2011feature,song2015diagnosis,infante2017use,you2017novel,pramono2019automatic,miranda2019comparative,solinski2020automatic}.
	These distinct features can be extracted by appropriate signal processing and mathematical transformations of the cough sounds.
	The features can then be used to train a sophisticated AI engine for performing the preliminary diagnosis solely based on cough.
	Our in-depth analysis of the pathomorphological alternations caused by COVID-19 in the respiratory system (reported in \cref{sec:section2}), shows that the alternations are distinct from those caused by other common non-COVID-19 respiratory diseases. This finding is corroborated by the meta-analysis of several recent independent studies (reported in \cref{sec:section2}) that show that COVID-19 infects the respiratory system in a distinct way.
	Therefore, it is logical to hypothesize that cough caused by COVID-19 is also likely to have distinct latent features and the risk of these features overlapping with those associated with other respiratory infections is low.
	These distinct latent features can be exploited to train a domain aware AI engine to differentiate COVID-19 cough from non-COVID-19 cough.
	Our experiments (\cref{t-sne}, \cref{sec:section3C}) show that this is indeed possible. 
	\item Cough manifests as a symptom in the majority (e.g., 67.7\% as per~\cite{world2020report}) but not all COVID-19 carriers.
	However, studies show that coughing is one of the key mechanisms for the social spreading of COVID-19~\cite{cough_spreader}.
	Droplets containing the virus emitted through cough  landing on  surfaces where the virus has been shown to survive for long periods of time has been reported as  the most prolific mechanism of spreading the COVID-19~\cite{NIH}.
	Hence, if a COVID-19 patient is not showing cough as a symptom, the patient is most likely not spreading as actively as a coughing COVID-19 patient.
	In other words, cough-based testing, even if far from being as sensitive as clinical testing, can actually directly help in reducing ${R_0}$~\cite{cascella2020features}.
	\item Due to the ease of measurement, a temperature scan is currently the predominant screening method for COVID-19, e.g., used at the airports.
	However, between cough and fever, the number of non-COVID-19 medical conditions that can cause fever are much larger than the non-COVID-19 conditions that can cause cough.
	Our analysis shows that cough contains COVID-19 specific features even if it is non-spontaneous, i.e., when the COVID-19 patient is asked to cough.
	This means cough can be used as a pre-screening method by asking the subject to simulate cough.
\end{enumerate}

\subsection{Contributions and paper contents}
\label{sec:section1C}

The contributions and contents of this paper are outlined below:
\begin{enumerate}[wide,font=\itshape]
	\item We analyze the pathomorphological changes caused by  COVID-19 in the respiratory system from the studies examining X-rays and CT scans of alive COVID-19 patients. 
	Our analysis also includes the autopsy report studies of deceased patients.
	The purpose of this analysis is to apply first principle-based approach.
	The goal is to see if the pathomorphological alterations caused by COVID-19 in the respiratory system (i.e., the part of body that produces a cough sound) are different from those caused by other common bacterial or viral infections.
	This is to determine if it is even theoretically possible for the COVID-19 cough to have any distinct latent features.
	The in-depth study of pertinent pathomorphological alterations suggests that it is possible. 
	\item Building on the insights from first principle-based approach and our prior work~\cite{bales2020can}, as well as several other independent studies~\cite{thorpe2001acoustic,chatrzarrin2011feature,song2015diagnosis,infante2017use,you2017novel,pramono2019automatic,miranda2019comparative,solinski2020automatic} that suggest  distinct latent features in cough sounds can be used for successful AI-based diagnosis of several respiratory diseases, we hypothesize that “Cough sound can be used at least for preliminary diagnosis of the COVID-19 by performing differential analysis of its unique latent features relative to other non-COVID-19 coughs. ”
	\item Continuing the medical literature review, we further identify and shortlist the non-COVID-19 respiratory syndromes that are relatively common and are known to cause similar-sounding cough as that of COVID-19 patients.
	The shortlist includes pertussis, bronchitis, influenza, asthma, pneumonia, bronchiolitis and croup. 
	\item Given that even the shortlist is too long to gather reliable data for this time sensitive project, we reduce the size of our data gathering campaign to a manageable one by leveraging the findings from literature which show that cough caused by the last five medical conditions in the shortlist above does have features unique to each condition.
	Therefore, in the interest of time, we go on to focus on the differential analysis of COVID-19 cough, and coughs associated with pertussis and bronchitis as these two conditions are not examined earlier. 
	\item We gather cough data of COVID-19, pertussis and bronchitis patients.
	Cough samples from COVID-19 patients include both spontaneous cough (symptomatic) and non-spontaneous (i.e., when the patient is asked to cough).
	This is to make the test applicable to those who may not be showing cough as a symptom yet but are already infected.
	We also gather cough samples from otherwise healthy individuals with no known medical condition, hereafter referred to as a normal cough.
	The normal cough is included in the analysis to see if it can be differentiated from the simulated cough produced by the COVID-19 patients.
	Using these data, we test the hypothesis using a variety of data analysis and pre-processing tools.
	Multiple alternative analysis approaches show that COVID-19 associated cough does have certain distinct features, at least when compared to pertussis, bronchitis and a normal cough. 
	\item Building on the insights from medical domain knowledge and cough data analysis, we develop an AI engine for preliminary diagnosis of COVID-19 from cough sounds.
	This engine runs on a cloud server with a front-end programmed as a simple user-friendly mobile app called AI4COVID-19.
	The app listens to cough when prompted, and then sends it to the AI engine wirelessly.
	The AI engine first runs a cough detection test to see if the recorded sound is a cough or not a cough.
	In case the sound is not a cough, it commands the app to indicate so. The cough detection part of the AI engine is designed to detect cough even in the presence of background noise. This is to make the app a useful screening tool even at public places such as airports and crowded shopping malls.
	If a cough is detected, it is passed on to the diagnosis part of the AI engine.  
	After the AI engine completes the analysis, the app renders the result with three possible outcomes:
	\begin{itemize} \setlength{\itemindent}{0.2in}
		\item COVID-19 likely.
		\item COVID-19 not likely.
		\item Test inconclusive.
	\end{itemize}   
	\item To make the results as reliable as possible with the limited data available at the moment, we propose and implement a risk-averse architecture for the AI engine. It consists of three parallel classification solutions designed independently by three teams. The classifiers' outcomes are consolidated by an automated mediator.  In the current design, each classifier has veto power, i.e., if all three classifiers do not agree, the app returns `Test inconclusive'. This architecture employs the "2\textsuperscript{nd} opinion" practice in medicine and reduces the rate of misdiagnosis, compared to stand alone classifiers with binary diagnosis, albeit at a cost of an increased rate of returning `Test inconclusive' result.

\end{enumerate}

%

\section{Methodology}
\subsection{Hypothesis Formulation and the Devising a Manageable Validation Strategy Guided by Relevant Clinical Findings}
\label{sec:section2}

Our hypothesis in question is: \textit{“Cough sounds of COVID-19 patients contain unique enough latent features to be used as a diagnosis medium”}.
In this section, we describe our first principle-based approach that established the theoretical possibility of our hypothesis to be true.
Then we describe the deep domain knowledge-based approach we take to reduce the amount of data required to test this hypothesis, thereby making this project feasible in a constrained time.

\subsubsection{Is COVID-19 cough unique enough to yield AI-based diagnosis?}
\label{sec:section2A}

Unfortunately, cough is a very common symptom of over a dozen medical conditions caused by either bacterial or viral respiratory infections not related to COVID-19~\cite{irwin2000diagnosis,chang2006cough,gibson2010cicada}.
Several non-respiratory conditions can also cause cough.
\Cref{cough-conditions} summarizes the non-COVID-19 medical conditions which are known to cause cough.
Theoretically, a cough based COVID-19 diagnosis, therefore, must take into account the cough sound data associated with all of the conditions listed in  \cref{cough-conditions}.

\begin{table}[t]
	\renewcommand\arraystretch{1.3}
	\caption{Non-COVID-19 Medical Conditions that can cause Cough}
	\centering
	\begin{tabular}{|>{\centering\arraybackslash} m{6cm}| >{\centering\arraybackslash} m{6cm}|} 
		\hline
		\bf RESPIRATORY &  \bf NON-RESPIRATORY \\ \hline
		Upper respiratory tract infection (mostly viral infections) & Gastro-esophageal reflux    \\ \hline
		Lower respiratory tract infection
		(pneumonia, bronchitis, bronchiolitis)
		&Drugs
		(angiotensin converting enzyme inhibitors; beta blockers)
		\\ \hline
		Upper airway cough syndrome  &	Laryngopharyngeal reflux  \\ \hline
		Pertussis, parapertussis&	Somatic cough syndrome 
		\\ \hline
		Tuberculosis &	Vocal cord dysfunction \\ \hline
		Asthma and allergies&	Obstructive sleep apnea \\ \hline
		Early interstitial fibrosis,
		cystic fibrosis &	Tic cough 
		\\ \hline
		Chronic obstructive pulmonary disease (emphysema, chronic bronchitis) &	Smoking 
		\\ \hline
		Postnasal drip 	&Foreign body
		\\ \hline
		Croup&	Mediastinal tumor
		\\ \hline
		Laryngitis &	Air pollutants
		\\ \hline
		Tracheitis	&Tracheo-esophageal fistula 
		\\ \hline
		Lung abscess &	Left-ventricular failure
		\\ \hline
		Lung tumor	&Congestive heart failure
		\\ \hline
		Pleural diseases&	Psychogenic cough 
		\\ \hline
		Interstitial lung disease&	Idiopathic cough 
		\\ \hline
	\end{tabular}
	\label{cough-conditions}
\end{table}

Trained physicians have been using cough sounds to perform a differential diagnosis among several respiratory conditions such as pneumonia, asthma, COPD, laryngitis and Tracheitis \cite{CR1, gibson2010cicada,CR3,CR4,CR5,CR6}. This is possible because in all these diseases the nature and location of the underlying irritant in the respiratory system is quite different leading to audibly distinct cough sounds.   However, an unaided human ear is not capable of differentiating coughs caused by the conditions listed in \cref{cough-conditions}.
Even with AI, in case there are no unique latent features in the cough sound of COVID-19 patients, there is a risk for a cough-based AI diagnosis tool to confuse the cough caused by any of the diseases identified in \cref{cough-conditions} with the cough caused by COVID-19.
A brute force-based approach to evaluate this risk would require gathering cough data from a large number of patients for each of the conditions listed in \cref{cough-conditions}.
This deluge of data can be then used to train a powerful AI engine, such as very deep neural network to see if it can differentiate COVID-19 cough from those caused by all of the other medical conditions listed in \cref{cough-conditions}.
This approach is not practical at the moment given that the gathering such all-encompassing data will take too much time, rendering this approach of no help for the current pandemic.

To ensure that our developed solution works in practice with useful accuracy while being trainable with timely available data, we take another approach that we call domain-aware AI-design.
Domain-aware here refers to the fact that the proposed AI engine does not solely rely on blind big data churning, e.g., through a deep neural network.
Instead it relies on the deep domain knowledge of medical researchers trained in respiratory and infectious diseases to assess and narrow down the hypothesis testing scope, and to minimize the amount of data needed to test our hypothesis. By deep domain knowledge of medical researchers, we mean the use of medical knowledge of medical experts in this field to analyze pathomorphological changes caused by COVID-19 in the respiratory system and thus to evaluate the feasibility of an AI-based approach using cough-based analysis. It also means identifying the location of irritant in different types of coughs and using that information for smart feature extraction and faster training.

To this end, the medical researchers in our team began with an in-depth analysis of the pathomorphological changes caused by COVID-19 in the respiratory system by examining the data reported in numerous recent X-rays and CT-scans based studies of COVID-19 patients.
The goal here is to see if the pathomorphological alterations caused by COVID-19 are distinct from that of other common medical conditions, particularly the ones identified in \cref{cough-conditions}, that are well known to cause cough.
If this turns out to be the case, then in cough caused by COVID-19 we should have latent features distinct from the cough caused by the other medical conditions.
An appropriately designed AI should then be able to pick these cough feature idiosyncratic to COVID-19 infection and yield a reliable diagnosis, given enough labeled data.
In the case of no such differences at pathomorphological level, the idea of cough based COVID-19 diagnosis should be dropped.
In that case, any AI-based diagnosis yielded from cough is more likely to be a frivolous correlation and not a meaningful causal relationship.
Such AI-based diagnosis will be an artifact of the training data rather than unique latent features of COVID-19 caused cough.
Such a domain oblivious solution irrespective of its performance in lab will not be useful in practice.

\subsubsection{Distinct pathomorphological alternations in respiratory system caused by COVID-19}
\label{sec:section2B}

In a recent study, it has been discovered that in COVID-19 infected people, there are distinct early pulmonary pathological signs even before the onset of the symptoms of COVID-19, such as dry cough, fever and some difficulty in breathing~\cite{tian2020pulmonary}.
Early histological changes include evident alveolar damage with alveolar edema and proteinaceous exudates in alveolar spaces, with granules; inflammatory clusters with fibrinoid material and multinucleated giant cells; vascular congestion.
Reactive alveolar epithelial hyperplasia and fibroblastic proliferation (fibroblast plugs) were indicative of early organization.

Contrary to the above observation of no early symptoms, it has also been noted that in some patients, COVID-19 leads to onset of pneumonia and pneumonia is marked by a peculiar cough~\cite{world2020report}.
However, pneumonia can also be caused by many other factors including non-COVID-19 viral or bacterial infections.
Therefore, the question arises: is there a difference between COVID-19 caused pneumonia and other types of pneumonia that can be expected to translate into a difference in associated cough’s latent features?
Recent study in ~\cite{bai2020performance} shows that compared to non-COVID-19 related pneumonia, COVID-19 related pneumonia on chest CT scan was more likely to have a peripheral distribution (80\% vs. 57\%), ground-glass opacity (91\% vs. 68\%), vascular thickening (59\% vs. 22\%), reverse halo sign (11\% vs. 9\%) and less likely to have a central+peripheral distribution (14\% vs. 35\%), air bronchogram (14\% vs. 23\%), pleural thickening (15\% vs. 33\%), pleural effusion (4\% vs. 39\%) and lymphadenopathy (2.7\% vs. 10.2\%). Hence, these findings clearly suggest that cough sound signatures with COVID-19 caused pneumonia are likely to have some idiosyncrasies stemming from the distinct underlying pathomorphological alterations. 

Moreover, CT scan-based studies also show that in the early stage of COVID-19 disease, it mainly manifests as an inflammatory infiltration restricted to the subpleural or peribronchovascular regions of one lung or both lungs, exhibiting patchy or segmental pure ground-glass opacities (GGOs) with vascular dilation.
There is an increasing range of pure GGOs and the involvement of multiple lobes of the lung, consolidation of lesions, and crazy-paving patterns during the progressive stage.
There are diffuse exudative lesions and lung "white-out" during an advanced stage~\cite{dai2020ct}. Furthermore, AI-based analyses of X-ray~\cite{wang2020covid,XRAY1,XRAY2,XRAY3} and CT scan~\cite{xu2020deep,CT2} of the respiratory system have also shown to exploit the differences in pathomorphological alternations caused by COVID-19 to perform differential diagnosis among bacterial infection, non-COVID-19 viral infection and COVID-19 viral infection, with good accuracy.
This further implies that COVID-19 affects the respiratory system in a fairly distinct way compared to other respiratory infections.
Therefore, it is logical to hypothesize and investigate that the sound waves of cough produced by the COVID-19 infected respiratory system may also have distinct latent features.

The feasibility of diagnosing several common respiratory diseases using cough is not only supported by prior studies~\cite{al2013signal,amrulloh2015cough,pramono2016cough} but also in a recent clinically validated and widely publicized study~\cite{porter2019prospective}.
In~\cite{porter2019prospective}, a large team of researchers showed that cough alone can be used to diagnose asthma, pneumonia, bronchiolitis, croup and lower respiratory tract infections with over 80\% sensitivity and specificity.

Recently, many machine learning teams around the world have started working on the idea of using cough sound for possible diagnosis of COVID-19, some interdependently and others inspired by our preliminary results in~\cite{bales2020can} and pre-print version of this work\footnote{\url{https://arxiv.org/pdf/2004.01275.pdf}}.
However, to the best of authors’ knowledge, this is the first work to propose and evaluate the feasibility of this idea, and develop and test the prototype of an AI  engine powered mobile app based solution for anytime, anywhere tele-testing and pre-screening for COVID-19.

%

\subsection{Data Description and Practical Viability of the Solution with Available Data }
\label{sec:section3}

As mentioned earlier, ideally cough data associated with all diseases listed in \cref{cough-conditions} is desirable for such a project.
However, gathering such mammoth data is not possible in this time-constrained project, as the COVID-19 pandemic needs rapid response.
To achieve meaningful results in the constrained time, we leverage domain knowledge, instead of just seeking big data.
From \cref{cough-conditions}, using the insights from \cref{sec:section2}, we shortlist cough causing infections that are most likely to confuse our AI engine due to similar pathomorphological changes in the respiratory system as of COVID-19 and, hence, similar cough signatures.
The shortlist includes pertussis, bronchitis, asthma, pneumonia, bronchiolitis, croup and influenza.
We further note that the prior study~\cite{porter2019prospective} has shown that cough associated with all of these seven medical conditions, except pertussis and bronchitis, have unique latent features.
We use findings from this earlier study to reduce the scope of our data gathering campaign and differential analysis to only the respiratory diseases, the cough for which has not been analyzed before for having unique features, i.e., pertussis and bronchitis.

\subsubsection{Data used for training cough detector}
\label{sec:section3A}

In order to make AI4COVID-19 app employable in a public place or where various background noises may exist (e.g., airport), we design and include a cough detector in our AI-Engine.
This cough detector acts as a filter before the diagnosis engine and is capable to distinguish cough sound from 50 types of common environmental noises.
To train and test this detector, we use the ESC-50 dataset~\cite{Piczak2015dataset} and the cough and non-cough sounds recorded from our own smartphone app.
The ESC-50 dataset is a publicly available dataset that provides a huge collection of human and environmental sounds.
This collection of sounds is categorized into 50 classes, one of these being cough sounds. We have used 1838 cough sounds and 3597 non-cough environmental sounds for training and testing of our cough detection system.

\subsubsection{Data used for training COVID-19 diagnosis engine}
\label{sec:section3B}

To train our cough diagnosis system, we collected cough samples from COVID-19 patients as well as pertussis and bronchitis patients. We also collected normal coughs, i.e., cough sounds from healthy people. At the time of writing, we had access to 96 bronchitis, 130 pertussis, 70 COVID-19, and 247 normal cough samples from different people, to train and test our diagnosis system. Obviously, these are very small numbers of samples and more data is needed to make the solution more generalizable. New COVID-19 cough samples are arriving daily, and we are using these unseen samples to test the trained algorithm. 

\subsubsection{Data pre-processing and visualization to evaluate the practical feasibility of AI4COVID-19}
\label{sec:section3C}
In \cref{sec:section2},  by applying medical domain knowledge, we analyzed the theoretical viability of our hypothesis. However, in AI-based solutions, theoretical viability does not guarantee practical viability as the end outcome depends on the quantity and quality of the data, in addition to the sophistication of the machine learning algorithm used.  Therefore, here we use the available cough data from the four classes, i.e., bronchitis, pertussis, COVID-19 and normal, to first evaluate the practical feasibility of a cough based COVID-19 diagnosis solution. 

All audio files used in our study are in uncompressed PCM 16-bit format with a sampling rate of 44.1 kHz and a fixed 3-second length. We convert the cough audio samples for all four classes into the Mel scale for further processing. The Mel scale is a pitch categorization where listeners judge changes in pitch to be equal in distance from one another along this scale. It is meant to make changes in frequency, such as with a spectrogram, more closely reflect audible changes. We used the Mel spectrogram over a typical frequency spectrogram because the Mel scale in the Mel spectrogram has unequal spacing in the frequency bands and provides a higher resolution (more informative) in lower frequencies and vice versa, as compared to equally spaced frequency bands in normal spectrogram \cite{mel-scale}. Since cough sounds are known to have more energy in lower frequencies therefore, the Mel spectrogram is a naturally suitable representation for cough sounds.
There are several methods for converting the frequency scale to the Mel. Here, we convert frequency $f$ into Mel scale m as:
\begin{equation} 
\label{eq:1} 
m=2595\times \log_{10}\left(1+\frac{f}{700}\right)
\end{equation}

We perform Cepstral analysis on the Mel spectrum of audio cough samples to compute their Cepstral coefficients, commonly known as \ac{MFCC} \cite{mfcc}. The extracted \ac{MFCC} features for every sample result in an $M \times N$ matrix, where each column represents one signal frame and each row represents extracted \ac{MFCC} features for a specific frame. The number of frames $N$ can vary from sample to sample. There are several possible ways to use these extracted features for classification. In our approach, we extract two $M \times 1$ \ac{MFCC} based feature vectors for each input cough sample and concatenate them into a single final $2M \times 1$ feature vector for that sample. For the first feature vector, we take the mean of \ac{MFCC} features corresponding to all the frames. For the second feature vector, we take the top $P$ $M \times 1$ \ac{PCA} projections \cite{pca} of the \ac{MFCC} features across all the frames and combine them into a single $M \times 1$ vector by taking their magnitude. Finally, we concatenate both feature vectors into a single $2M \times 1$ feature vector. This approach is further illustrated in \cref{fig:CML-MC} in \cref{sec:section4C}.

\begin{figure}[t]
	\centering
	\includegraphics[width=0.4\columnwidth]{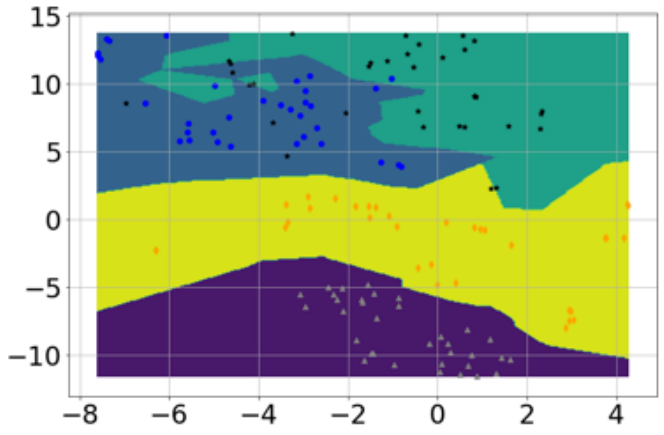}
	\caption{Visualization of features for the four classes via t-SNE (gray triangles correspond to  normal, blue circles correspond to bronchitis, black stars correspond to pertussis and orange diamonds represent COVID-19 cough.}
	\label{t-sne}
\end{figure}

Since the features extracted from cough audio are multi-dimensional, in order to visualize the features, a nonlinear dimensionality reduction technique, t-distributed Stochastic Neighbor Embedding (t-SNE) \cite{tsne} is applied, as it is well-suited for embedding high-dimensional data in a low-dimensional space of two-dimensions. In particular, this technique models each high-dimensional object by a two-dimensional point such that similar objects are modeled by nearby points and dissimilar objects are modeled by distant points with high probability. This visualization allows us to interpret the features in the form of clusters or classes with classification decision boundaries. \Cref{t-sne} illustrates the 2-D visualization of these features for the four classes through t-SNE with classification decision boundaries/contours.
It can be observed from the figure that different cough types possess features distinct from each other, and the features for COVID-19 are different from other cough types, such as bronchitis and pertussis.
Hence, this observation suggests the practical viability of AI-powered cough based preliminary diagnosis for COVID-19 encouraging us to proceed towards an AI-engine design for maximum accuracy and efficient implementation to enable app-based deployment.

%

\subsection{The AI4COVID-19 AI-Engine}
\label{sec:section4}

In this section we explain the system architecture and the details of a two-stage solution that we developed for: 1)~detection of cough sound from mixed cough, non-cough and noisy sounds;  and 2)~diagnosis of COVID-19 from the cough sound. 

The training data is used to train different variants of deep learning and one classical machine learning algorithm as described in this section. After these models are trained, the pre-trained models for both cough detection and COVID-19 diagnosis are then implemented at the cloud server. The app then provides a user interface for using these pre-trained models. Another advantage of cloud-based implementation is the possibility of refining the model continuously as more data becomes available, as no update in the app is required for the refinement in the back-end AI-based diagnosis engine.
\subsubsection{System architecture}
\label{sec:section4A}

\begin{figure*}
	\centering
    \includegraphics[width=\columnwidth]{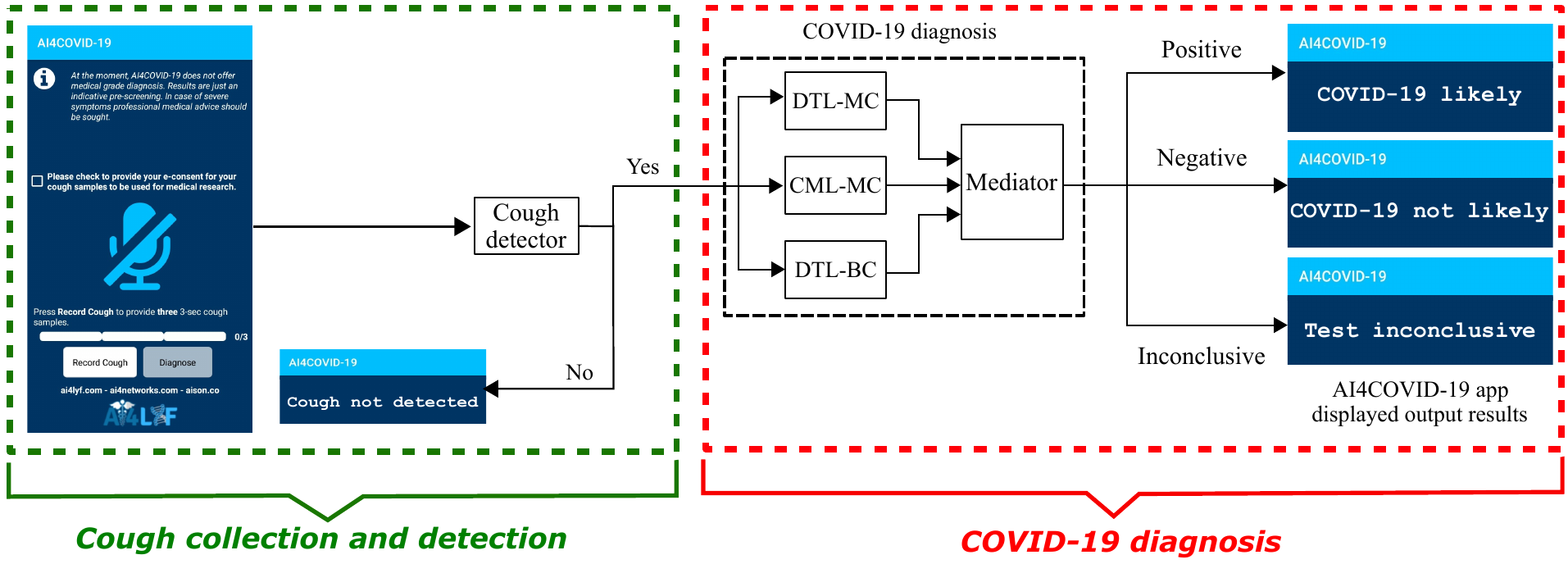}
	\caption{Proposed system architecture and flow diagram of AI4COVID-19, showing snapshot of Smartphone App at user front-end and back-end cloud AI-engine blocks consisting of Cough Detector block (further elaborated in \cref{fig:detection} and \cref{sec:section4B}) and COVID-19 diagnosis block containing Deep Transfer Learning-based Multi-Class classifier (DTL-MC), Classical Machine Learning-based Multi-Class classifier (CML-MC) and Deep Transfer Learning-based Binary-Class classifier (DTL-BC)  (further elaborated in \cref{fig:CML-MC} and \cref{sec:section4C}).}
	\label{overall_system}
\end{figure*}

\begin{figure}
	\centering
	\includegraphics[width=0.75\columnwidth]{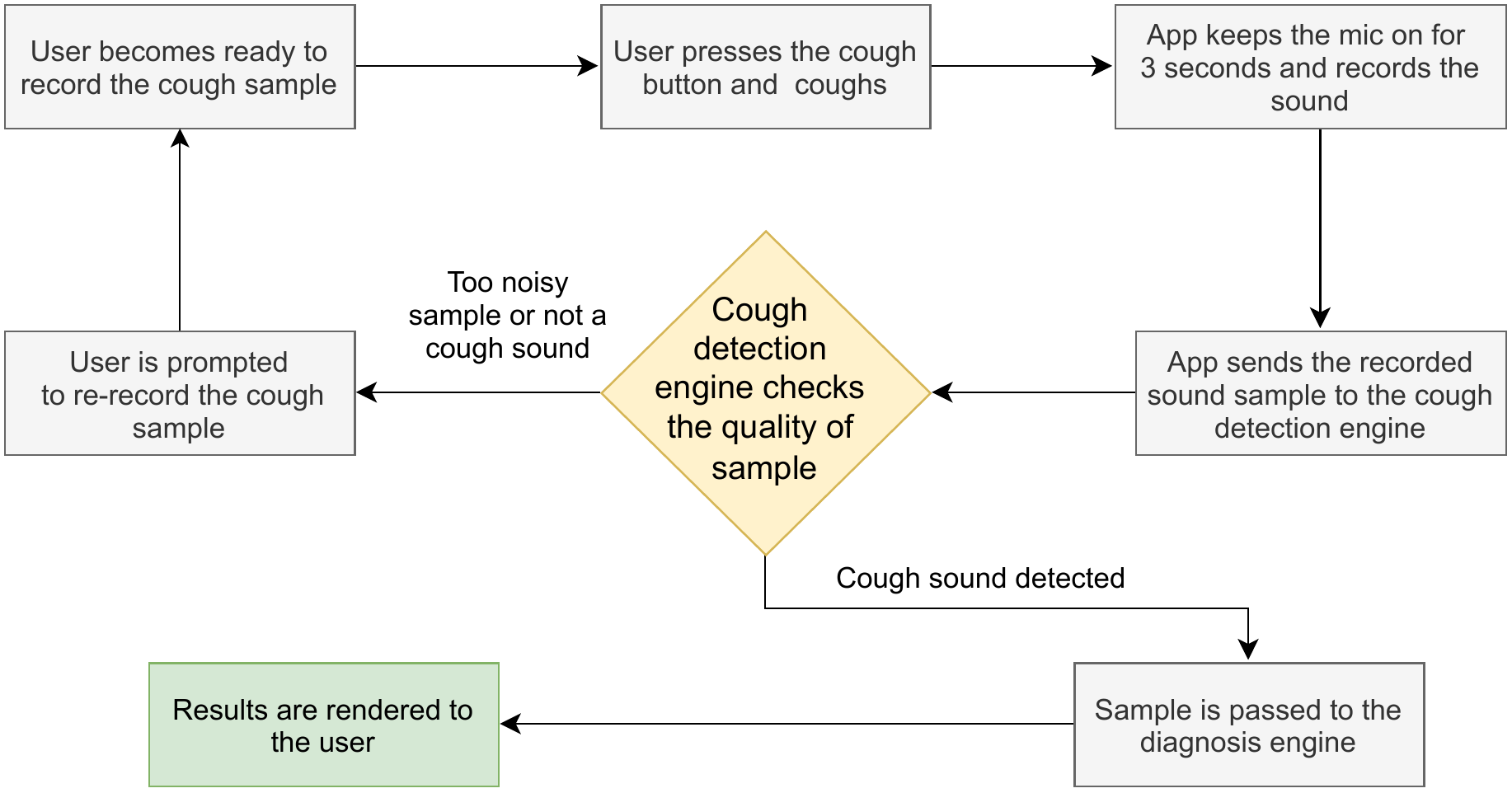}
	\caption{A flow chart highlighting the steps of the proposed system.}
	\label{flow_chart}
\end{figure}

The overall system architecture is illustrated in \cref{overall_system} and a flow chart highlighting the complete steps is shown in \cref{flow_chart}.
The smartphone app records sound/cough when prompted by the press and release button. The recorded sounds are forwarded to the server when the diagnosis button is pressed.  At the server, the sounds are first fed into the cough detector. In case, the sound is not detected as cough, the server commands the app to prompt so. In case, the sound is detected as a cough, the sound is forwarded to three parallel, different classifier systems, i.e., Deep Transfer Learning-based Multi Class classifier (DTL-MC), Classical Machine Learning-based Multi Class classifier (CML-MC) and Deep Transfer Learning-based Binary Class classifier (DTL-BC). The results of all these three classifiers are then passed on to a mediator. The app reports a diagnosis only if all three classifiers return identical classification results. If the classifiers do not agree, the app returns `test inconclusive'. This tri-pronged mediator centered architecture is designed to  minimize the probability of misdiagnosis. With this architecture, results show that AI4COVID-19 engine predicting `COVID-19 likely' when the subject is not suffering from COVID-19 or vice-versa is extremely low when validated on the testing data available at the time of writing. The multi-pronged architecture is inspired by the "second opinion" practice in health care. The added caution here is that the three (diagnosis) opinions are solicited, each with veto power. How this architecture manages to reduce the overall misdiagnosis rate of the AI4COVID-19 despite the relatively higher misdiagnoses rate of individual classifiers is further explained in \cref{sec:section5C} through \eqref{med1} and \eqref{med2}.

For app implementation in real-time, to ensure stricter quality control, we plan to run these pre-trained algorithms on at least three cough samples from the same patient and then make a preliminary diagnosis based on majority voting. Also, the cough detector, implemented before COVID-19 diagnosis (see \cref{overall_system} and \cref{flow_chart}) is ensuring some quality control by passing only those cough samples to the COVID-19 diagnosis engine that are of satisfactory quality. If samples are of poor quality, for example, a lot of background noise or the sound is too low, it rejects those samples by not detecting them as cough and therefore, not passing them on for diagnosis. In this case, the user is prompted to re-record the cough sample.

\begin{figure*}
	\centering
	\includegraphics[width=\columnwidth]{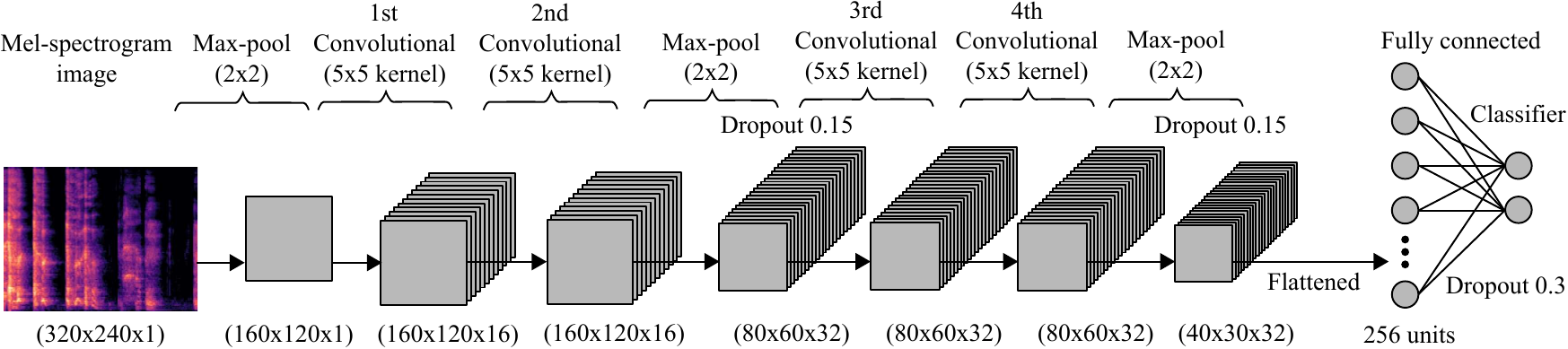}
	\caption{Cough detection classifier.}
	\label{fig:detection}
\end{figure*}

The details of detection and diagnosis classifiers are presented below.

\subsubsection{Cough detection}
\label{sec:section4B}

The recorded cough sample is forwarded to our cloud-based server where the cough detector engine first computes its Mel-spectrogram (as explained in \cref{sec:section3C}) with 128 Mel-components (bands). This image is then resized and converted into grayscale to unify the intensity scaling and reduce the image dimensions, resulting in a $320 \times 240 \times 1$ dimensional image. The resultant image is then fed into our Convolutional Neural Network (CNN) based classifier to decide whether the recorded input sound is of cough or not.


An overview of our used CNN structure is shown in \cref{fig:detection}. As the input Mel spectrogram image is of high dimensions, it’s first passed through a $2\times2$ max-pooling layer to reduce the overall model complexity before proceeding. This is followed by two blocks of layers, each block comprising two convolutional layers followed by a $2\times2$ max pooling layer and a 0.15 dropout. Convolutional layers in first block use 16 filters and a $5\times5$ kernel size, whereas the second block uses 32 filters each in both convolutional layers. The learned complex features from these 4 convolutional layers are flattened and then passed to a fully connected layer of 256 neurons followed by a 0.30 dropout layer to prevent overfitting. Finally, the output layer with 2 neurons and a softmax activation function is used to classify between cough and not cough for the given input.  ReLU is used as the activation function for all convolutional layers in this model, while Adam~\cite{kingma2014adam} is used as the optimizer due to its relatively better efficiency and flexibility. A binary cross entropy loss function completes the detection model.



\subsubsection{COVID-19 diagnosis}
\label{sec:section4C}

When the input sound is detected to be cough by the cough detection engine, it is forwarded to our tri-pronged mediator-centered AI engine to diagnose between COVID-19 and non-COVID-19 coughs. In order to produce results with maximum reliability, with the limited data available at the moment, the three classifiers used in the system use different approaches and are designed independently by three teams and cross-validated \cite{kfold}.

The three classification approaches are described below.

\paragraph{Deep Transfer Learning-based Multi Class classifier (DTL-MC)}
The first solution leverages a CNN-based four class classifier, using Mel spectrograms (described above) as input. The four classes here are cough caused by 1)~COVID-19, 2)~pertussis, 3)~bronchitis or 4)~normal person with no known respiratory infection. Similar CNN architecture used for cough detection is used here with a slight modification to make it a four class classifier (instead of binary classifier previously) by changing the number of neurons in output layer to four neurons for classifying the input between four possible output classes. Deep transfer learning \cite{pan2009survey} is used here to transfer the knowledge (features) learned by cough detection model (trained using relatively more data) to the similar diagnosis model. This allowed us to train a deep architecture (see \cref{fig:detection}) using limited amount of training data, as the basic features of the input Mel-spectrogram characterizing cough are already learned and only  fine-tuning is required to learn more subtle features using new disease data. In this DTL-MC model, we froze the initial weights of the first convolutional layer, as the initial layers learn low-level latent features, and only allowed other layers to fine-tune their weights. This transfer learning approach allowed us to get better performance (reported in \cref{sec:section5}) than training on disease data from scratch.

\begin{figure}[t]
	\centering
	\includegraphics[width=0.5\columnwidth]{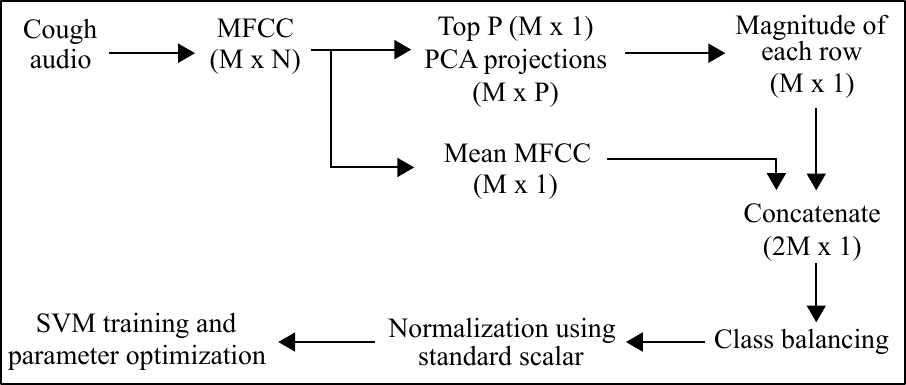}
	\caption{Classical Machine Learning-based Multi-Class classifier (CML-MC).}
	\label{fig:CML-MC}
\end{figure}

\paragraph{{Classical Machine Learning-based Multi Class classifier (CML-MC)}}

A second parallel diagnosis test uses classic machine learning instead of deep learning. This to mitigate the over-fitting that may still be happening in the deep learning-based classifier due to small amount of training data.  To maximize independence among the classifiers that together constitute the AI diagnosis engine, the 2\textsuperscript{nd} classifier begins with a different pre-processing of cough sounds. Instead of using a spectrogram like the first classifier, it uses MFCC and PCA based feature extraction as explained in \cref{sec:section3C}. These smart features are then fed into a multi-class support vector machine (SVM) for classification. Class balance is achieved by sampling from each class randomly such that the number of samples equals to the number of minority class samples, i.e., class with the lowest number of samples.
Using the concatenated feature matrix (of mean MFCC and top few PCAs) as input, we perform SVM with k-fold validation for 100,000 iterations. This approach is illustrated in \cref{fig:CML-MC}.

\paragraph{{Deep Transfer Learning-based Binary Class classifier (DTL-BC)}} 
The third parallel diagnosis test also uses deep transfer learning based CNN on the Mel spectrogram image of the input cough samples, similar to the first branch of the AI engine, but performs only binary classification of the same input, i.e., is the cough associated COVID-19 or not. The CNN structure used for this technique is similar to the one used for the cough detector (see \cref{fig:detection}).

%

\begin{figure}
	\centering
	\includegraphics[width=0.25\columnwidth]{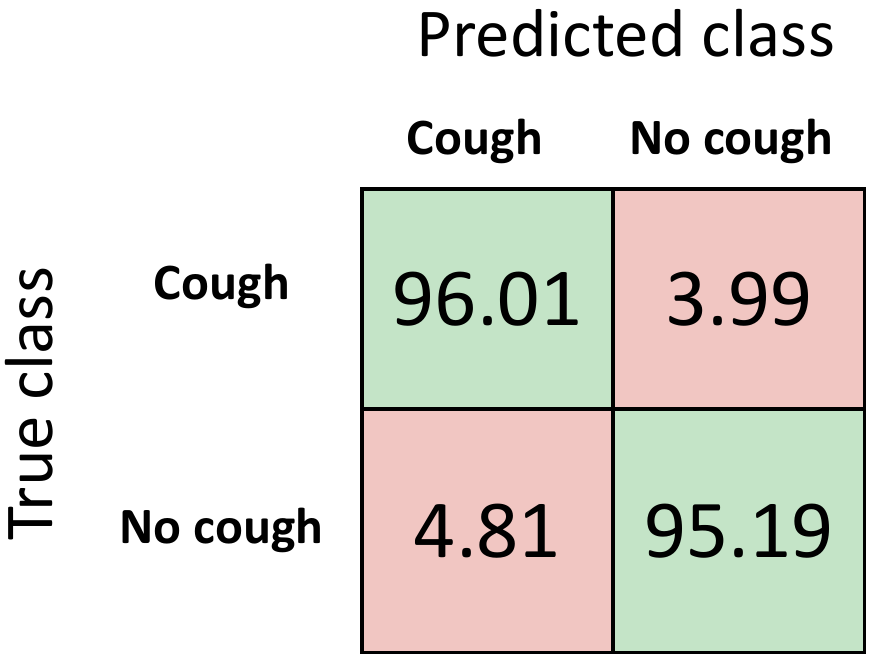}
	\caption{Normalized mean confusion matrix for cough detection (in percentage) using 5-fold cross validation.}
	\label{detection_CF}
\end{figure}

\begin{table}
	\renewcommand\arraystretch{1.2}
	\caption{Performance Metrics for Cough Detection}
	\centering
	\begin{tabular}
		{|P{1.2cm}|P{1.4cm}|P{1.4cm}|P{1.4cm}|P{1.2cm}|} 
		\hline
		\bf  F1-Score
		(\%)  & \bf Sensitivity
		(\%)
		& \bf Specificity
		(\%)
		& \bf  Precision
		(\%)
		& \bf Accuracy
		(\%)
		\\
		\hline
			95.61&96.01&95.19&95.22&95.60 \\
		\hline
	\end{tabular}
	\label{detection_PM}
\end{table}

\section{Results}
\label{sec:section5}

In order to evaluate the model we use the performance metrics of \textit{accuracy}, \textit{specificity}, \textit{sensitivity/recall}, \textit{precision}, \textit{F1}-score on validation set and also cross-validate the models. The \textit{accuracy} here refers to the overall accuracy of the model. We use k-fold cross validation methodology, that is well-suited to evaluate the performance of machine learning models on limited data \cite{kfold}.
These performance metrics are based on mean confusion
matrices from cross-validation. In addition, we have used regularization techniques to prevent the problem of over-fitting, for example, we tune the regularization parameter of SVM against the cross-validation accuracy and choose those parameters that gave us the best generalizability of the models. Tuning of the various hyper-parameters (number of hidden layers, learning rate, activation functions, dropout rate) of deep neural network-based models has also been performed, based on the cross-validation accuracy. Furthermore, the decay of model loss versus the number of epochs has been investigated to rule out the possibility of over-fitting.






\subsection{Cough detection}
\label{sec:section5A}

The confusion matrix and performance metrics for detection algorithm are reported in \cref{detection_CF} and \cref{detection_PM}, respectively.
Results demonstrate that our cough detection algorithm can classify between cough event and no cough event with an overall accuracy of 95.60\%.

\begin{figure}
	\centering
	\includegraphics[width=0.6\columnwidth]{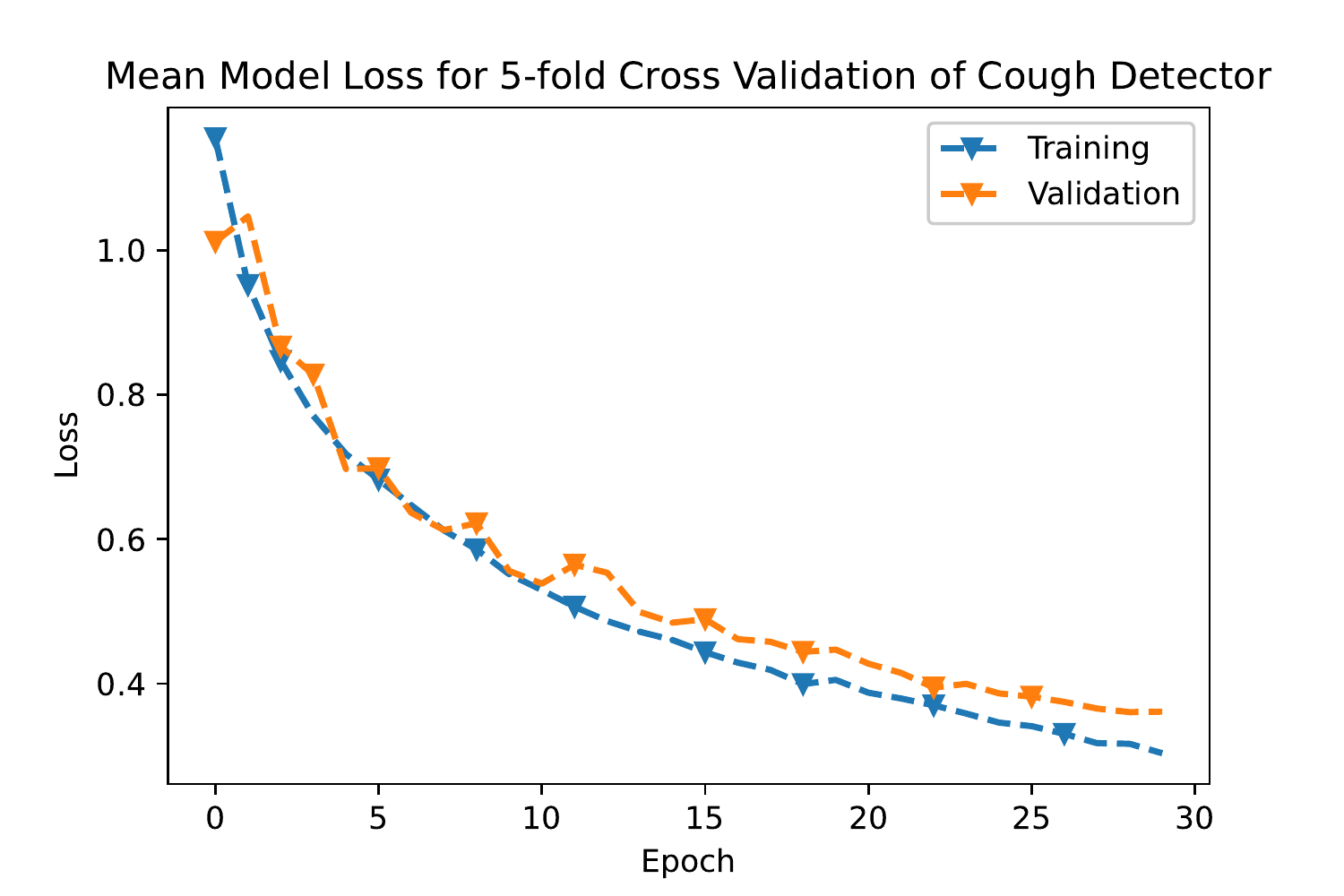}
	\caption{Mean model loss for 5-fold cross validation of cough detector.}
	\label{Error_Detection}
\end{figure}

The error graph of mean loss versus epochs of this neural network based model, for both training and validation data sets is shown in \cref{Error_Detection}. The decay of both the training and testing curve shows that this model has not been over-fitted.

\subsection{COVID-19 diagnosis}
\label{sec:section5B}

\begin{figure}
	\centering
	\includegraphics[width=0.3\columnwidth]{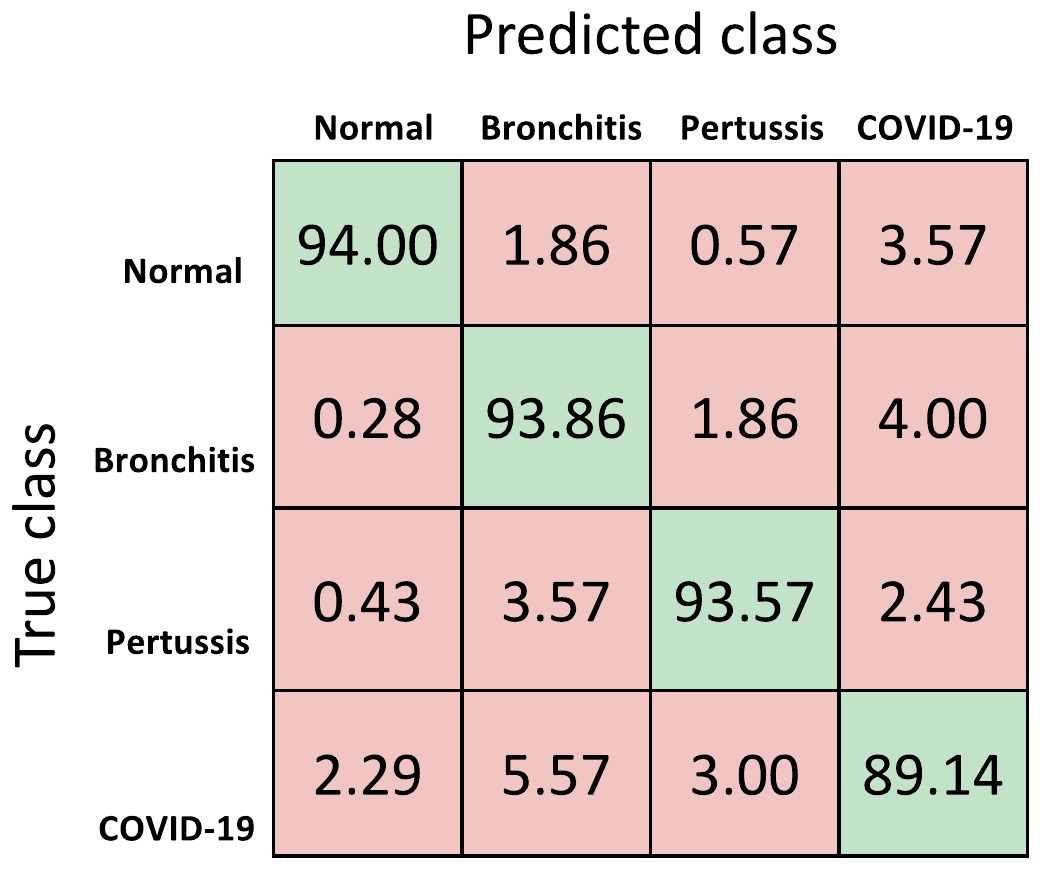}
	\caption{Normalized mean confusion matrix for cough diagnosis (in percentage) for DTL-MC using 5-fold cross validation.}
	\label{CM_DL-MC}
\end{figure}

\begin{figure}
	\centering
	\includegraphics[width=0.6\columnwidth]{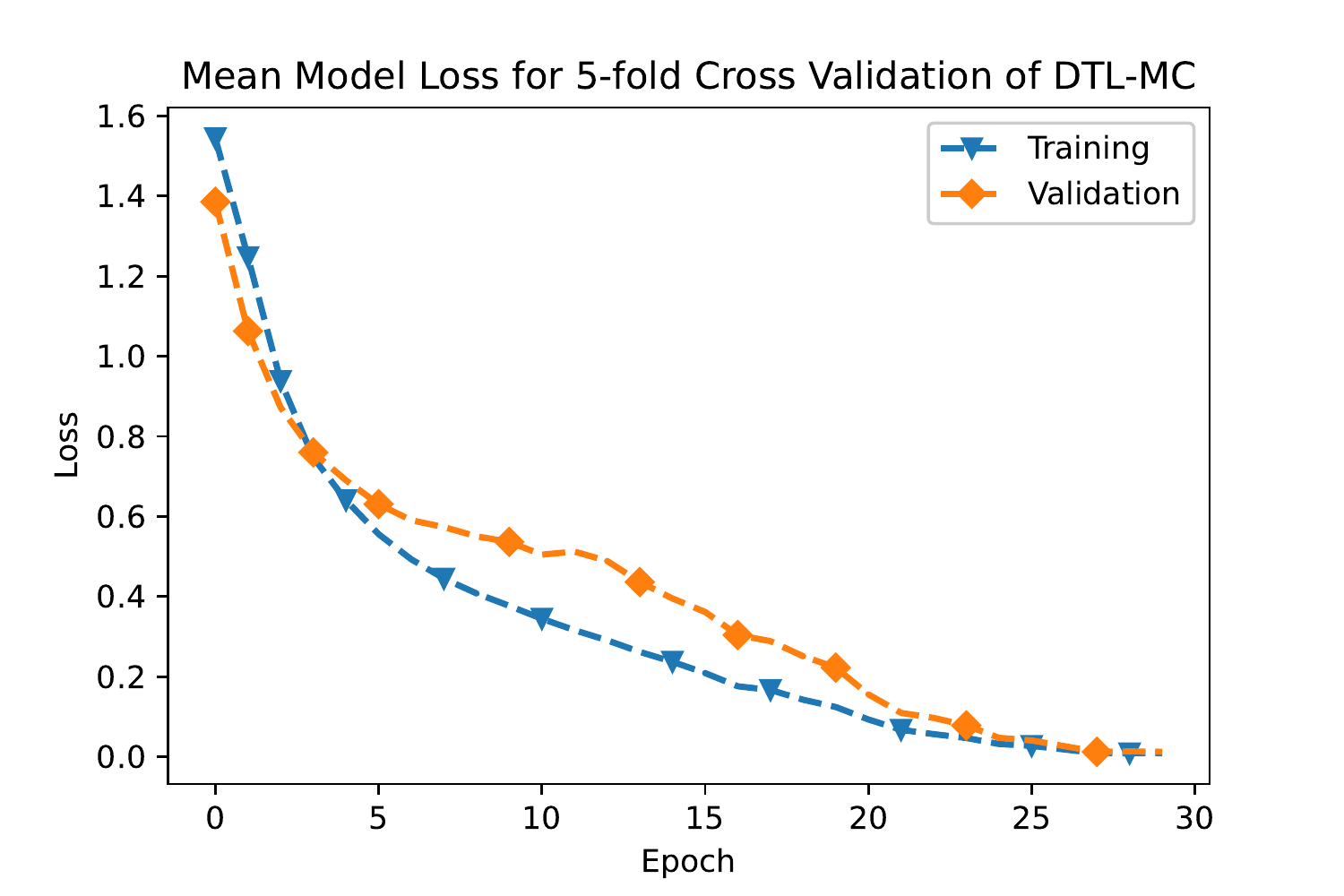}
	\caption{Mean model loss for 5-fold cross validation of DTL-MC.}
	\label{Error_Detection_DTL_MC}
\end{figure}

The performance metrics for the first classifier, that is DTL-MC classifier are reported in \cref{DL-MC-PM}.
At the moment, with limited data available, the overall accuracy of deep transfer learning based multi-class classifier is 92.64\%. The mean normalized confusion matrix resulting from this approach is shown in \cref{CM_DL-MC}. 
Future work will continue to improve this model as more training data becomes available for CNN.   \cref{Error_Detection_DTL_MC} shows the mean loss versus epochs of the DTL-MC classifier, for both training and validation data sets. Both the curves start to saturate after around 25 epochs, indicating a reasonable learning time, without over-fitting.

\begin{table}
	\renewcommand\arraystretch{1.2}
	\caption{Performance Metrics for DTL-MC}
	\centering
	\begin{tabular}{|l |P{1.3cm }| P{1.4cm}|P{1.2cm}|P{1.1cm}|P{1.1cm}|} 
		\hline
		&	\bf  F1-Score
		(\%)  & \bf Sensitivity
		(\%)
		& \bf Specificity
		(\%)
		& \bf  Precision
		(\%)
		& \bf Accuracy
		(\%)
		\\
		\hline
		Overall&-&-&-&-& 92.64\\ \hline
		COVID-19 &	89.52&89.14& 96.67&	89.91&	-\\
		\hline
		Pertussis&	94.04&	93.57&	98.19&	94.51	&-\\\hline
		Bronchitis&	91.63&	93.86&	96.33	&89.50	&-\\ \hline
		Normal&		95.43	&94.00&	99.00&	96.90&	-\\ 
		\hline
	\end{tabular}
	\label{DL-MC-PM}
\end{table}

\begin{figure}
	\centering
	\includegraphics[width=0.3\columnwidth]{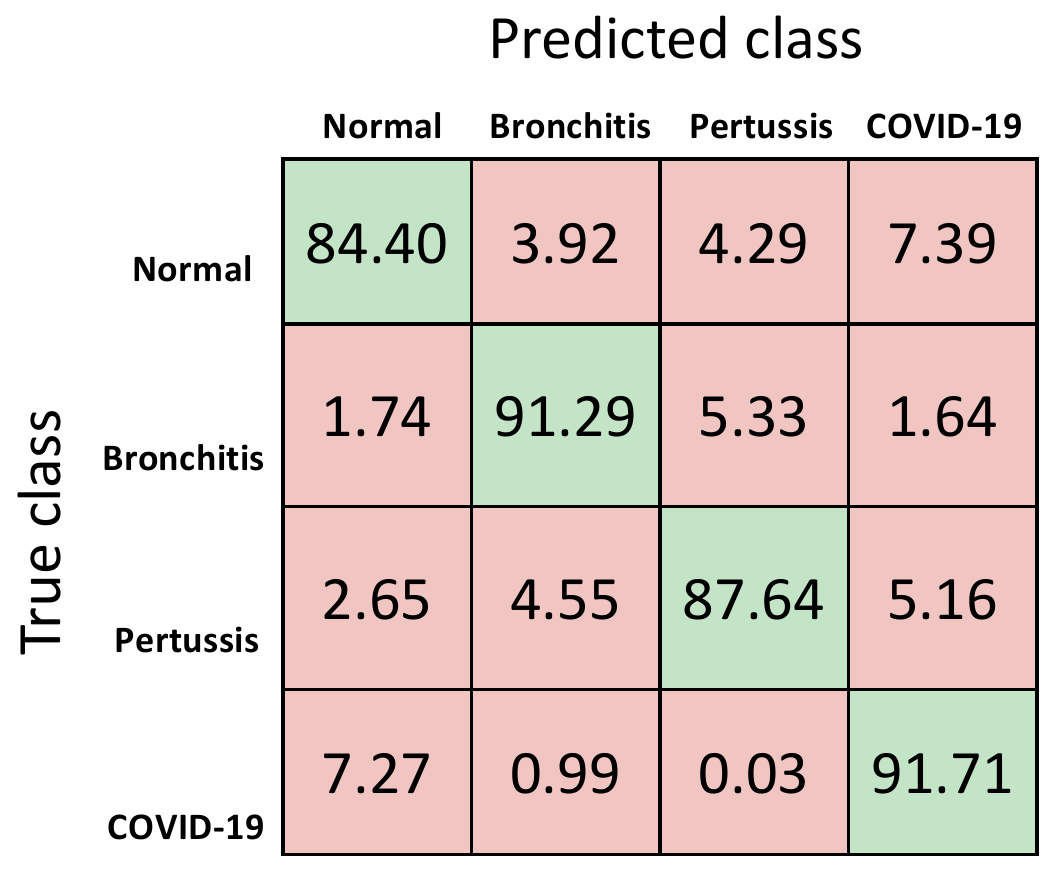}
	\caption{Normalized mean confusion matrix for cough diagnosis (in percentage) for CML-MC using 5-fold cross validation.}
	\label{CML-MC-CM}
\end{figure}

\begin{table}
\centering
	\renewcommand\arraystretch{1.2}
	\caption{Performance Metrics for CML-MC}
	\begin{tabular}{|l |P{1.3cm }| P{1.4cm}|P{1.2cm}|P{1.1cm}|P{1.1cm}|} 
		\hline
		&	\bf  F1-Score
		(\%)  & \bf Sensitivity
		(\%)
		& \bf Specificity
		(\%)
		& \bf  Precision
		(\%)
		& \bf Accuracy
		(\%)
		\\
		\hline
		Overall&-&	-&	-&	-&	88.76\\ \hline
		COVID-19 &89.08	&91.71	&95.27&	86.60	&-\\
		\hline
		Pertussis&	88.84	&87.64&	96.78	&90.08&	-\\\hline
		Bronchitis&	90.94&	91.29&	96.84&	90.61&	-\\ \hline
		Normal&		86.09&	84.40&	96.11&	87.86&	-\\ 
		\hline
	\end{tabular}
	\label{CML-MC-PM}
\end{table}

\begin{figure}
	\centering
	\captionsetup{justification=centering}
	\includegraphics[width=0.5\columnwidth]{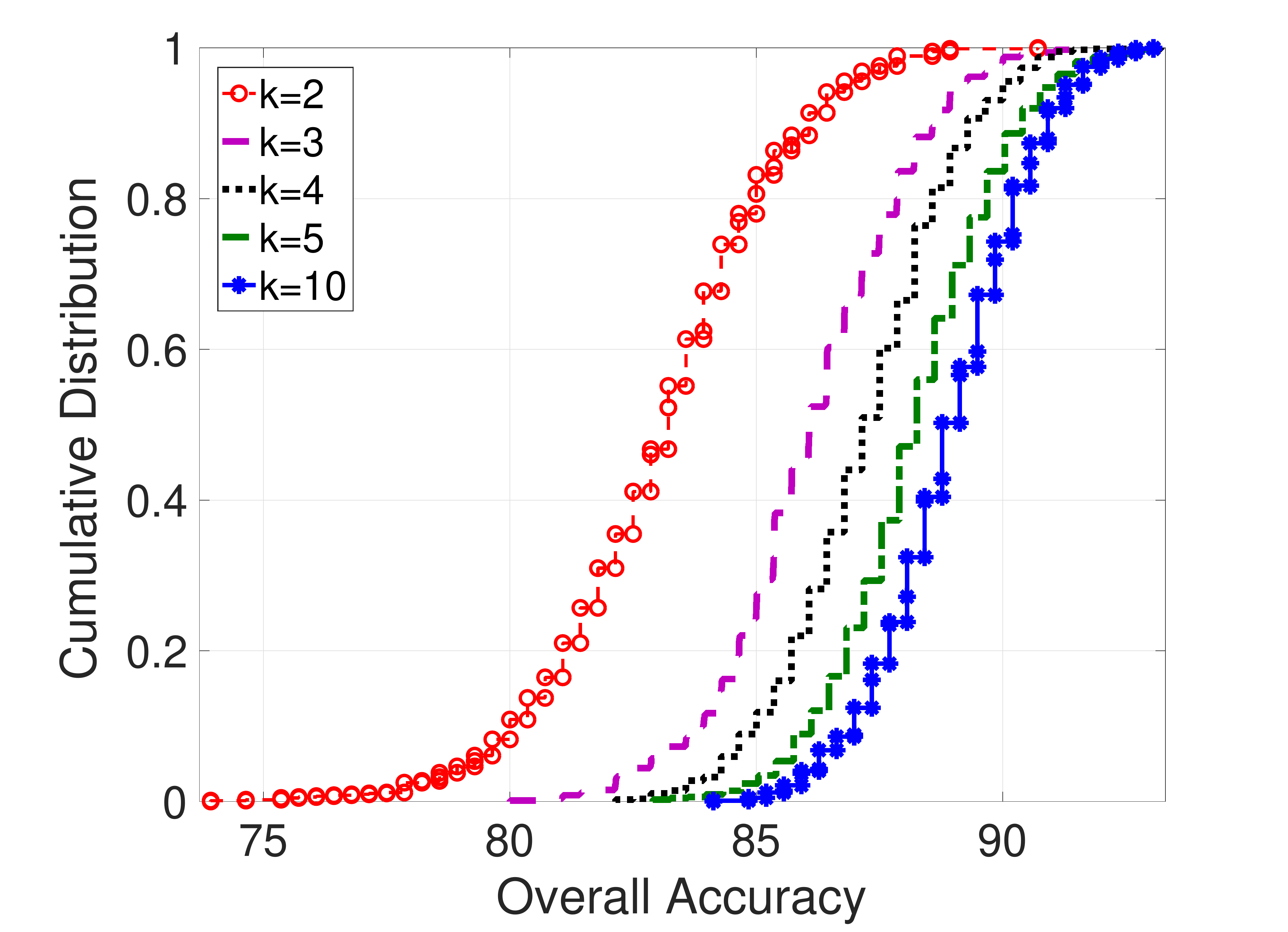}
	\caption{Overall accuracy CDF for varying k-fold experiments in CML-MC approach.}
	\label{CDFs}
\end{figure}

For the second classifier, i.e., CML-MC classifier, the normalized mean confusion using 5-fold cross validation is shown in \cref{CML-MC-CM} and the CDF of overall accuracy with varying k’s in k-fold cross validation is shown in \cref{CDFs}.
\cref{CML-MC-PM} reports the performance metrics for this approach, utilizing data available at this moment. Results indicate an overall accuracy of 88.76\%.

\begin{table}
\centering
	\renewcommand\arraystretch{1.2}
	\caption{Performance Metrics for DTL-BC}
	\begin{tabular}
		{|P{1.2cm}|P{1.4cm}|P{1.4cm}|P{1.4cm}|P{1.2cm}|} 
		\hline
		\bf  F1-Score
		(\%)  & \bf Sensitivity
		(\%)
		& \bf Specificity
		(\%)
		& \bf  Precision
		(\%)
		& \bf Accuracy
		(\%)
		\\
		\hline
		92.97&94.57&91.14&91.43&92.85\\
		\hline
	\end{tabular}
	\label{DL-BC-PM}
\end{table}

\begin{figure}
	\centering
	\includegraphics[width=0.3\columnwidth]{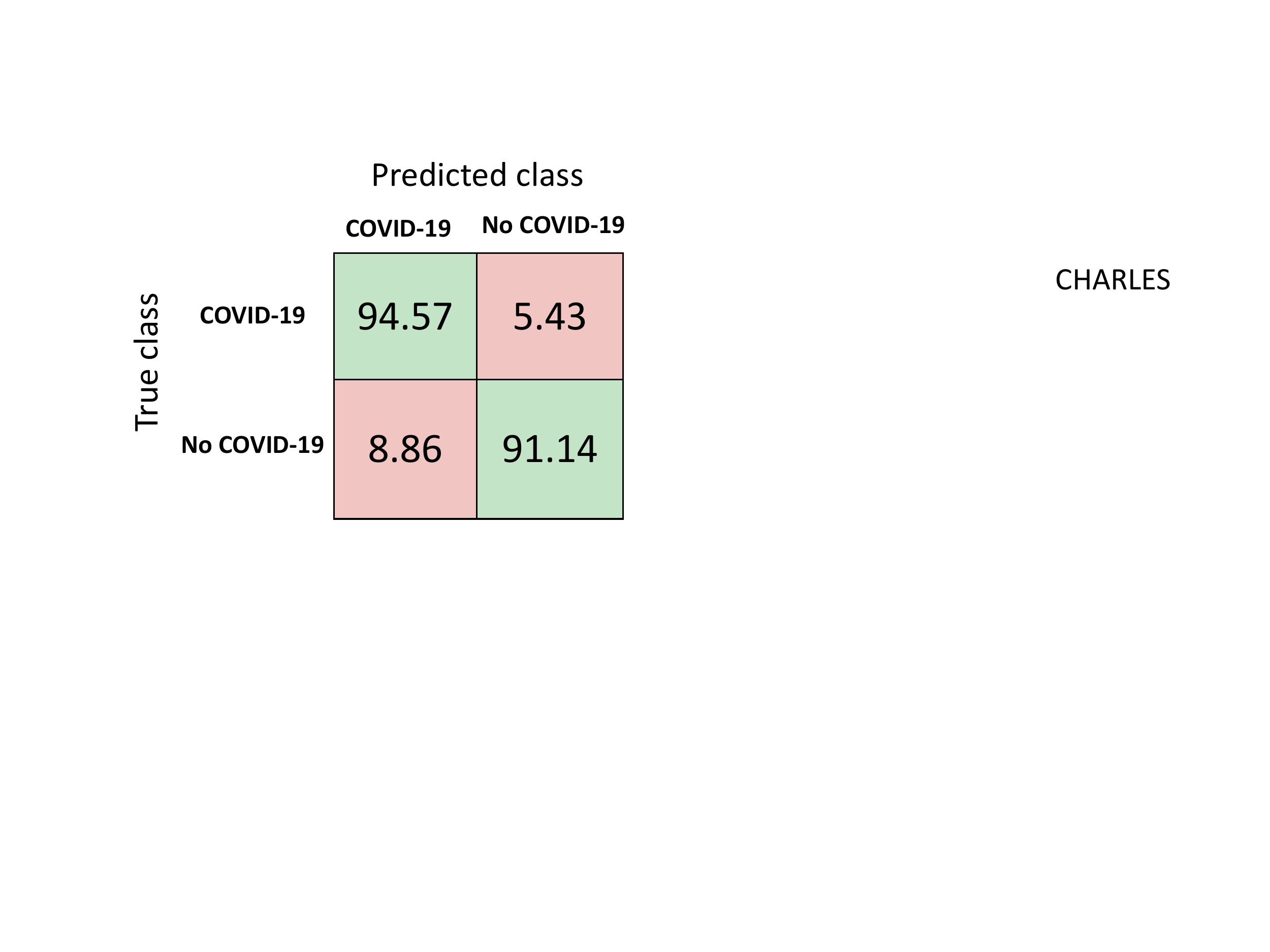}
	\caption{Normalized mean confusion matrix for cough diagnosis (in percentage) for DTL-BC using 5-fold cross validation.}
	\label{CM_DL-BC}
\end{figure}

\begin{figure}
	\centering
	\includegraphics[width=0.6\columnwidth]{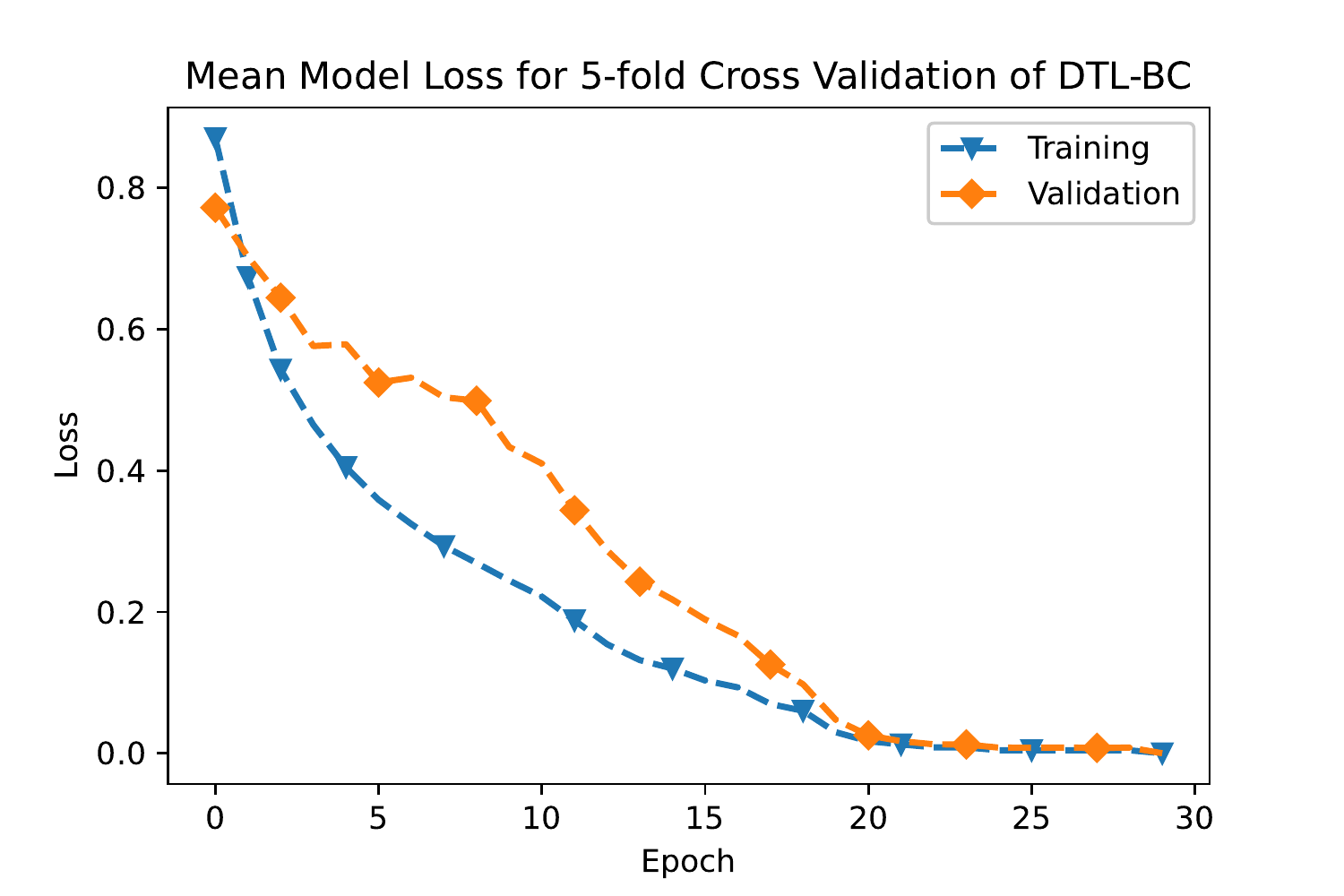}
	\caption{Mean model loss for 5-fold cross validation of DTL-BC.}
	\label{Error_Detection_DTL_BC}
\end{figure}

Performance metrics for the third approach, that is DTL-BC are reported in \cref{DL-BC-PM}, with the normalized mean confusion matrix shown in \cref{CM_DL-BC}. The classification accuracy with this approach is 92.85\%.
The loss versus number of epochs for both training and validation is illustrated in \cref{Error_Detection_DTL_BC}. Here, both the curves start to level off after 20 epochs, hence depicting a reasonable training time, while avoiding over-fitting.
Currently, the number of non-COVID cough samples are much larger than COVID-19 cough samples when binary classification is chosen. Once more data becomes available, the current classification accuracy using DTL-BC is likely to increase.

The performance of the two deep learning-based classifiers (DTL-MC and DTL-BC) is superior than the manual feature extraction based classic machine learning classifier (CML-MC).
This is expected, because with shortage of training data circumvented via transfer learning,  considerable amount of training data and automatic feature extraction capability of the deep neural network are expected to extract even more subtle distinct features hidden in the data than the manual feature extraction used in the second classifier, i.e., CML-MC.   

\subsection{Overall performance under independence assumption}
\label{sec:section5C}

After analyzing the performance of the three different classifiers, we now analyze the overall performance of AI4COVID-19 AI engine that utilizes a mediator-based architecture. This architecture will yield optimal performance when its prongs (i.e., the classifiers) are fully independent. 

The independence of the three classifiers depends on dependence among the training data fed into these classifiers, as well as the similarity among the classifier's internal architectures.   
Therefore, in reality, the classifiers will never be truly independent, because of following key reasons: (i)~Even if we use unique training data for each classifier, there will be some dependence (e.g., correlation introduced by age group, gender, native language etc). (ii)~Even if we manage to choose fully independent training data for each classifier, the similarities in the architectures of classifiers would introduce some degree of dependence.

However, lacking absolute independence does not completely eliminate the advantages of proposed multi-pronged architecture. 
This is similar to a scenario when a second diagnosis sought from a physician, who has the same speciality, reads same medical literature, and has correlated neuroanatomy as the first physician, is considered an independent opinion for all practical purposes and is known to reduce misdiagnosis rates, though in strict theoretical sense it is not fully independent diagnosis.  

 Acknowledging that the three classifiers are not fully independent but they will become almost independent by using unique training data when more COVID-19 cough data becomes available, in the following, we analyze the performance of overall AI architecture under the independence assumption. This is to compare the misdiagnosis rate of individual classifier decisions versus the mediator's decision.

Let $k_1$, $k_2$, $k_3$ be the predicted class labels for the three classifiers, DTL-MC, CML-MC and DTL-BC, respectively and $k_f$ be the predicted diagnosis result of the app. The possible values that $k_f$ can take are `COVID-19 likely' ($C$), `COVID-19 not likely' ($C'$) and `test inconclusive' ($I$).
Then, the probability that the app predicts `COVID-19 likely', when the patient actually has COVID-19, can be calculated as:

\begin{equation}
P (k_f=C|C) = P(k_1=C|C)\cdot P(k_2=C|C) \cdot P(k_3=C|C)  \nonumber      \\
\hspace{-4mm} = 0.891 \cdot 0.917 \cdot 0.946 =	  0.773   
\end{equation} 
The probability that the app predicts `COVID not likely' when the subject actually does not have COVID-19 can be represented as:

\begin{equation}
P (k_f=C'|C') = P(k_1=C'|C')\cdot P(k_2=C'|C') \cdot P(k_3=C'|C') \nonumber \\
= 0.966 \cdot 0.952 \cdot 0.911 =	  0.838
\end{equation} 
The app can also predict `COVID-19 likely' when the subject is not suffering from COVID-19 or vice-versa. In these cases, we can write the probabilities as:

\begin{equation}
P (k_f=C|C') = P(k_1=C|C')\cdot P(k_2=C|C') \cdot P(k_3=C|C') \nonumber\\
\hspace{7mm} = 0.033\cdot 0.047 \cdot 0.088=	1.365 \times 10^{-4}  \label{med1}
\end{equation}

\begin{equation}
P (k_f=C'|C) = P(k_1=C'|C)\cdot P(k_2=C'|C) \cdot P(k_3=C'|C) \nonumber\\
\hspace{7mm} = 0.108 \cdot 0.082\cdot 0.054=4.782 \times 10^{-4} \label{med2}
\end{equation}  

Equations \eqref{med1} and \eqref{med2} signify the importance of the mediator in our proposed architecture and show how this risk-averse architecture is able to reduce the overall misdiagnosis rate of  AI4COVID-19. From  \eqref{med1} and  \eqref{med2}, both the false negative as well as the false positive rate of the overall architecture are near zero.   Note that none of the classifiers have near zero misdiagnosis rate simultaneously for both healthy and COVID-19 cases. For a given classifier low \ac{FPR} is at the cost of high \ac{FNR} and vice versa. Mediator counters the over sensitivity or under sensitivity of the individual classifiers by masking it with the `Test inconclusive' result. i.e., from \eqref{med1}, the lowest false positive rate of DTL-MC classifier is the most contributing factor in  the near-zero probability the app will predict `COVID-19 likely' when the subject is not suffering from COVID-19.
The most contributing factor in the near-zero probability that the app will predict `COVID-19 not likely' when the subject is actually suffering from COVID-19, is the lowest false negative rate of DTL-BC classifier, as observed from \eqref{med2}.
In other words, the mediator in AI4COVID-19 architecture complements the weakness of one classifier with the strength of other and vice versa,  resulting in reduced misdiagnosis rate as compared to using these classifiers independently, i.e., without the proposed mediator.

In the cases where the reports `Test inconclusive', the test subject can either have COVID-19 or not, in reality. The respective probabilities for those cases are:


\begin{equation}
P (k_f=I|C) =  1- [ P(k_f=C|C) + P(k_f=C'|C) ] \nonumber \\ 
\hspace{25mm}  1- [0.773+ 4.782 \times 10^{-4}] = 0.226
\end{equation}
\begin{equation}
P (k_f=I|C') =    1- [ P(k_f=C|C') + P(k_f=C'|C') ] \nonumber \\ 
\hspace{17mm}  1- [1.365 \times 10^{-4} + 0.838] =  0.161
\end{equation}

Currently, the app would predict an inconclusive test result 38.7\% of the time ($P(k_f=I) = P (k_f=I|C)+P (k_f=I|C')$). This percentage can be reduced by switching to a mediation scheme where app result reflects simple or weighted majority of the $N$ number of classifiers. This scheme will be explored once more data becomes available. The results are summarized in \cref{overall_table}.
The numbers here just indicate how including the proposed mediator-based architecture may reduce the misdiagnosis rate compared to using individual classifiers. These probabilities are under independence assumption and can change depending on the degree of dependence between the training data and architectures of the individual classifiers, as explained earlier. We can capture this dependency factor by introducing a co-efficient, $d_i$ in each of the above six calculated probabilities, where  $i= 1 \dots 6$. The values of $d_i$'s can be estimated empirically once more data becomes available in the future and can in turn be used to determine the weights to be assigned to each classifier in weighted average based mediator design.

\begin{table*}[t]
	\renewcommand\arraystretch{1.2}
	\caption{The Overall Current Performance of AI4COVID-19 AI Engine}
	\centering
	\begin{tabular}{ |c|c| } 
		\hline
		\bf Event & \bf Probability \\
		\hline
		App reports `COVID-19 likely' when the subject actually has COVID-19& $0.773\hspace{0.3mm} d_1$  \\ 
		\hline
		App reports `COVID-19 likely' when the subject actually does not have COVID-19 & $1.365 \times 10^{-4} \hspace{0.3mm}d_2$  \\  \hline
		App reports `COVID-19 not likely' when the subject  actually does not have COVID-19 & $0.838 \hspace{0.3mm}d_3$ \\ 
		\hline
		App reports `COVID-19 not likely' when the subject  actually has  COVID-19 & $ 4.782 \times 10^{-4} \hspace{0.3mm}d_4$\\ 
		\hline
		App reports `test inconclusive' when the subject  actually has  COVID-19 & $0.226\hspace{0.3mm}d_5$ \\ 
		\hline
		App reports `test inconclusive' when the subject  actually does not have COVID-19 & $0.161 \hspace{0.3mm}d_6$\\ 
		\hline
	\end{tabular}
	\label{overall_table}
\end{table*}

%

\section{Discussion}
\subsection{Potential Utilities of the AI4COVID-19}
\label{sec:section6}

The AI4COVID-19 app based on preliminary diagnosis is not meant to replace or compete with the medical grade testing by any means. Instead, the proposed solution offers the following complementing use cases to control the pandemic.
\begin{enumerate}[wide,font=\itshape]
	\item Enabling tele-screening for anyone, anywhere, anytime.
	\item Addressing the shortage of testing facilities. This is particularly useful in remote areas of the world where medics have no option but to rely on phone based or questioner based tele-screening. In such places, the app can act as a clinical decision assistance tool.  
	\item Opportunity to protect medics from unnecessary exposure, particularly for non-critical patients where the medical advice for whom anyway would be “stay at home” or “self-isolate” to wait for self-healing.
	\item Minimizing covert spread that happens to be the biggest problem . 
	\item Tracing and monitoring the spread. This is particularly easy with AI4COVID-19 as the cough samples can be spatio-temporally tagged anonymously, without having to compromise the patient’s privacy.
	\item AI4COVID-19 can be used as a low cost screening tool, instead of or in addition to the temperature scanner at the airports, borders or elsewhere as needed. This is possible because our tests show that the app can diagnose COVID-19 even in a non-spontaneous cough of COVID-19 positive people. 
	The cost of using such an app-based solution would be significantly low, since it can be readily installed on any existing smartphone using the existing internet connections, by a large number of people simultaneously.
	\item The app can help in enabling and maintaining informed social distancing and self-isolation.
	\item By default the app can provide centralized record of tests with spatial and temporal stamps.
	Thus, the data gathered from the app can be used for long term planning of medical care and policy making.
\end{enumerate}

%

\subsection{Comparison and contrast of AI4COVID-19 with existing studies}

Existing methods to screen COVID-19 patients include Nucleic Acid Amplification Tests (NAAT), such as real-time Reverse Transcription Polymerase Chain Reaction (rRT-PCR).
While far more sensitive than proposed method, these methods are marked by limitations identified in Section \ref{sec:section1A} that includes limited geographical and temporal availability, high cost, large turnaround time, requirement of in-person visits to hospitals or mobile labs and the need and shortage of protective equipment.  
In contrast, AI4COVID-19 is useable anywhere, anytime for anyone. 

 Recent AI-based studies towards COVID-19 preliminary diagnosis include the use of either X-ray~\cite{wang2020covid,XRAY1,XRAY2,XRAY3} or CT Scan~\cite{xu2020deep,CT2,CT3}. These methods demonstrate comparable or higher sensitivities, ranging from 72\% to 96\%, compared to proposed approach.    However,  both of these approaches still require a visit to a well-equipped clinical facilities and does not meet the utilities identified in Section  \ref{sec:section6}. In contrast, AI4COVID-19 is the only screening method proposed in the literature so far that can be used in-situ and eliminates the need for an in-person visit to the testing facility or getting out of homes or places of self-isolation, thereby meeting all use cases identified in Section \ref{sec:section6}.



\subsection{Key Limitations of Current Version of AI4COVID-19}
\label{sec:section7}

At the time of writing, the performance of AI4COVID-19 app is limited by the following factors:
\begin{enumerate}[wide,font=\itshape]
	\item The quantity of the training and testing data.
	Due to time constraints and difficulty of getting cough data, we could gather data only from a small number of patients for each of the four groups.
	We tried to minimize the impact of this limitation by combining data hungry approaches that are capable of extracting more hidden features i.e., deep learning, with the ML approaches that can work with a small amount of data through manual feature extraction.  The shortage of training data was also to some extent circumvented by using transfer learning in the deep learning based classifiers.
	Still, the need for more data cannot be overemphasized. 
	\item The quality of the training and testing data: We have strived to ensure that the data is correctly labeled.
	However, any error in the labeling of the data that managed to slip through our scrutiny is likely to impact reported performance.
	Such impact can be particularly pronounced when the data is not that big in the first place.
	\item Our in-depth medical differential analysis suggested that COVID-19 associated pathomorphological alternations are fairly distinct, and hence cough of COVID-19 patients is likely to have at least some distinct latent features.
	However, this does not guarantee the absence of overlap in COVID-19 cough features and those of diseases not included in the training and testing. The approach we used to combat this issue is the clever mediator-based architecture that practically eliminates misdiagnosis by declaring test to be inconclusive if the cough samples are even slightly confusing i.e., lying very close to decision boundaries. Still, we are working to address this limitation in future releases of AI4COVID-19 by incorporating cough associated with other non-COVID-19 medical conditions identified in \cref{cough-conditions} as well as including other dimensions such as age, gender, smoking or non-smoking status and certain bio markers.
	\item Large scale trial-based validation to test the generalization capability: In the end, the only way to evaluate the generalization capability and practical performance of the proposed AI4COVID-19 based testing is a large scale medically supervised validation in real world. 
	The findings of this paper provide promising enough preliminary results and proof of concept to encourage first systematic large-scale cough data gathering campaigns followed by large scale trials.  Once the testing of the prototype app on a much larger data set is completed, the provision of automatic updates will also be enabled.

\item	In the current prototype design, all AI processing happens at the cloud. The app is just a thin client that records and sends the audio data to the server where the AI engine resides. Due to low complexity, the app does not have stringent CPU and RAM requirements and it can run on most smartphones. This cloud-based design allows the screening to be done not only via commodity smartphones but also via a web portal link accessible in any browser. In the future, to enable offline screening using an edge device such as smart phone, we plan to investigate edge-based implementation of the modified lightweight version of the proposed AI-engine. This will be done by edge AI techniques such as distilled deep leering. The potential of distilled deep learning for enabling edge device based medical diagnoses has been verified in our recent work \cite{morghan}.

\end{enumerate}
	
\subsection{Planned Future upgrades of AI4COVID-19}
AI4COVID-19 accuracy can be improved by incorporating other acoustic data such as breathing sound and speech. Moreover, for higher accuracy and better generalization across larger populations, we also plan to investigate the impact of incorporating meta-data such as age, gender, smoking, non-smoking, ethnicity and medical history. The accuracy is also likely to improve by including multi-sensory data instead of relying on only acoustic data and meta-data.  For example, recent studies show that in a small fraction of COVID-19 patients, cutaneous anomalies are part of the symptoms \cite{cutaneous}. Therefore, including skin images in addition to acoustic data may help improve the diagnosis.  Another planned upgrade is the inclusion of bio-markers that can be measured by wearable sensors such as wristbands, rings and skin patches or ambient sensors such as infrared cameras or wireless sensors, which can also lead to more reliable results. The examples of bio-markers that are worthy of investigation that can be easily collected via aforementioned wearable or ambient sensors include respiration rate, temperature, blood oxygen saturation, pulse rate,  heart rate variability, resting heart rate, blood pressure, mean arterial pressure, stroke volume, sweat level, systematic vesicular resistance, cardiac output, pulse pressure and cardiac index.

%

\section{Conclusion}
\label{sec:section8}

Scarcity, cost and long turnaround time of clinical testing are key factors behind covert rapid spread of the COVID-19 pandemic. Motivated by the urgent need, this paper presents a ubiquitously deployable AI-based preliminary diagnosis tool for COVID-19 using cough sound via a mobile app.  The core idea of the tool is inspired by our independent prior studies that show  cough can be used as a test medium for diagnosis of a variety of respiratory diseases using AI. To see if this idea is extendable to COVID-19, we perform in-depth differential analysis of the pathomorphological alternations caused by COVID-19 relative to other cough causing medical conditions. We note that the way COVID-19 affects the respiratory system is substantially unique and hence, cough associated with it is likely to have unique latent features as well. We validate the idea further by the visualization of latent features in cough of COVID-19 patients and two common infections, pertussis and bronchitis as well as non-infectious coughs. Building on the insights from the medical domain knowledge, we propose and develop a tri-pronged mediator centered AI-engine for the cough-based diagnosis of COVID-19, named AI4COVID-19. The results show that the AI4COVID-19 app is able to diagnose  COVID-19 with negligible misdiagnosis probability thanks to its risk-avert architecture.

Despite its impressive performance, AI4COVID-19 is not meant to compete with clinical testing. Instead, it offers a unique functional tool for timely, cost-effective and most importantly safe monitoring, tracing, tracking and thus, controlling the rampant spread of the global pandemic by virtually enabling testing for everyone.  While we are working on improving the AI4COVID-19, this paper is meant to present a proof of concept to encourage community support for more labeled data followed by large scale trials. We hope that the AI4COVID-19 app can be leveraged to pre-screen for COVID-19 at a population scale, particularly in regions around the world where the pandemic is spreading covertly due to the lack of testing. The AI4COVID-19 enabled tele-screening can alleviate the crushing burden on the overwhelmed medical systems around the world and  help save countless lives.

%

\section*{Acknowledgment}
This work is dedicated to those affected by the COVID-19 pandemic and those who are helping to fight this battle in anyway they can.

\bibliography{IEEEabrv,references}
\bibliographystyle{IEEEtran}

\end{document}